\documentclass[iop]{emulateapj}

\usepackage{txfonts}
\usepackage{graphicx}
\usepackage{graphicx}
\usepackage{natbib}
\newcommand{\ks}{\mbox{km~s$^{-1}$}}

\usepackage{natbib}

\begin{document}
\title{Kelvin-Helmholtz instability in solar chromospheric jets: theory and observation}
\vskip1.0truecm
\author{
D. Kuridze$^{1,3}$, T. V. Zaqarashvili$^{2,3,4}$,  V. Henriques$^{1}$, M. Mathioudakis$^{1}$, F. P. Keenan${^1}$, and A. Hanslmeier$^{2}$}
\affil
{$^1$Astrophysics Research Centre, School of Mathematics and Physics, Queen's University, Belfast, BT7~1NN, Northern Ireland, UK; e-mail: d.kuridze@qub.ac.uk} %;\
\affil{$^2$IGAM, Institute of Physics, University of Graz, Universit\"atsplatz 5, 8010 Graz, Austria}
\affil{$^3$Abastumani Astrophysical Observatory at Ilia State University, 3/5 Cholokashvili avenue, 0162 Tbilisi, Georgia}
\affil{$^4$Space Research Institute, Austrian Academy of Sciences, Schmiedlstrasse 6, 8042 Graz, Austria}
\date{received / accepted }

%%%%%%%%%%%%%%%%%%%%%%%%%%%%%%%%%%%%%%%%%%%%%%%%%%%%%%%%%%%%%%%%%%%%%%%%%%%%%%%%%%%%%%%%%%%%%%%%%%%%%%%%

\begin{abstract}

Using data obtained by the high resolution CRisp Imaging SpectroPolarimeter instrument on the Swedish 1-m Solar Telescope,
we investigate the dynamics and stability of quiet-Sun chromospheric jets observed at disk center.  
Small-scale features, such as Rapid Redshifted and Blueshifted Excursions,
appearing as high speed jets in the wings of the H$\alpha$ line, are characterized by short 
lifetimes and rapid fading 
without any descending behavior.
To study the theoretical aspects of their stability without considering their formation mechanism,  
we model chromospheric jets as twisted magnetic flux tubes moving along their axis, and use 
the ideal linear incompressible magnetohydrodynamic approximation to derive the governing dispersion equation. 
Analytical solutions of the dispersion equation indicate that this type of jet is unstable to Kelvin-Helmholtz instability (KHI),
with a very short (few seconds) instability growth time at high upflow speeds.  
The generated vortices 
and unresolved turbulent flows associated with the KHI
could be observed as broadening of chromospheric spectral lines.  
Analysis of the H$\alpha$ line profiles shows that the detected structures have enhanced line widths with respect to the background.
We also investigate the stability of a larger scale H$\alpha$ jet that was ejected along the line-of-sight. 
Vortex-like features, rapidly developing around the jet's boundary, are considered as evidence of the KHI.
The analysis of the energy equation in the partially ionized plasma shows that the ion-neutral collisions may lead to the fast 
heating of the KH vortices
over timescales comparable to the lifetime of chromospheric jets.
 
\end{abstract}

%%%%%%%%%%%%%%%%%%%%%%%%%%%%%%%%%%%%%%%%%%%%%%%%%%%%%%%%%%%%%%%%%%%%%%%%%%%%%%%%%%%%%%%%%%%%%%%%%%%%%%%%

\section{Introduction}

The chromosphere is a highly inhomogeneous layer of the solar atmosphere, 
populated by a wide range of dynamical features such as spicules (type I and II), fibrils, mottles, rapid excursions (REs) which can be redshifted (RRE) or blueshifted (RBE), 
and surges \citep[see the review by][]{2012SSRv..169..181T}.
These are small-scale, short-lived, jet-like plasma structures  
observed near the network boundaries ubiquitously between the photosphere and the corona.
However, they have different physical properties and evolution cycles. In particular,
the traditional (type I) spicules and mottles have lifetimes ranging from 1 - 12 minutes and are characterized by rising motions with speeds of $\sim$20 - 40 km s$^{-1}$. 
At the last stage of their lifetime, type I spicules either 
fall back to the low chromosphere with a speed comparable to their upflow value, or fade gradually without any descending motion. 
\cite{2012ApJ...750...16Z} have reported that around $60\%$ of type I spicules have a complete cycle of ascent and descent movement.
There are larger scale H$\alpha$ jets in the chromosphere such as 
macrospicules and surges which could have diameter and flow speeds of to a few Mms and $\mathrm{\sim100~km~s^{-1}}$, respectively. 

A very prominent class of spicular type jets are the so-called type II spicules and their on-disk counterparts, 
RBEs/RREs \citep[][hereafter termed REs]{2007PASJ...59S.655D, 2008ApJ...679L.167L}.  
REs are absorption features detected in the blue and red wings of chromospheric spectral lines \citep{2008ApJ...679L.167L, 2009ApJ...705..272R,2013ApJ...769...44S,2015ApJ...802...26K}.  
They are slender $(\mathrm{\sim200~km~in~width)}$, short-lived $(\mathrm{\sim40~s)}$ with higher apparent velocities of $\mathrm{\sim50-150~km~s^{-1}}$. 

The formation mechanism of spicules is not well understood, with several different ones proposed, including: magnetic reconnection \citep{1969PASJ...21..128U,1990SoPh..128..333P}, granular buffeting \citep{1979SoPh...61...23R}, velocity pulses \citep{1982SoPh...75...99S}, 
rebound shocks \citep{1982ApJ...257..345H,2010A&A...519A...8M}, Alfv\'en waves \citep{1981SoPh...70...25H}, and p-mode leakage \citep{2004Natur.430..536D}. 
These models can explain the formation of classical spicules to some extent, but the formation of REs/Type II spicules 
with their extremely high upflow speeds ($\sim50 - 150~\ks$) remains a mystery. 
\cite{2011JGRA..116.9104S,2014ApJ...796L..23S} found that the high heating rate in the strong magnetic field region 
in the upper chromosphere is able to produce type-II spicules, based 
on their model of the strong damping of Alfv\'en waves via plasma-neutral collisions.
MHD simulations also struggle to reproduce the observational characteristics of spicules, including heights, lifetimes, velocities, densities and temperatures.

REs are also characterized by rapid fading in the chromospheric spectral line without any descending behavior.
It is suggested that the rapid disappearance of the spicular jets may be the result of their fast 
heating to transition region (TR) temperatures. \cite{2011Sci...331...55D} have provided evidence that  
TR and coronal brightenings in AIA passbands are occurring co-spatially and co-temporally with chromospheric RBEs. 
 Furthermore, \cite{2011A&A...532L...1M} showed that the coronal hole large spicules observed with HINODE in the Ca {\sc{ii}} H line 
are appearing in the TR O {\sc{v}} 629.76 {\AA} line observed with the SUMER instrument on-board SOHO.
The analysis of the differential emission measure distribution for a region dominated by spicules also indicates that they are heated to transition region temperatures \citep{2012SoPh..280..425V}. 
Recently, \cite{2014ApJ...792L..15P} studied the thermal evolution of type II spicules using combine observations with the HINODE and Interface Region Imaging Spectrograph (IRIS) satellites.  
They showed that the fading of spicules from the chromospheric Ca {\sc{ii}} H line
is caused by rapid heating of the upward moving spicular plasma to higher temperatures. 
More recently, \cite{2015ApJ...799L...3R} found spectral signatures for REs in the IRIS C {\sc{ii}} 1335
and 1336 {\AA} and Si iv 1394 and 1403 {\AA} spectral lines, and interpreted those as evidence that REs are heated
to at least TR temperatures. Furthermore, the IRIS observations revealed the prevalence of
fast network jets with lifetimes of 20 to 80 seconds, widths of $\leqslant$300 km 
and upflow speeds of 80-250 km~s$^{-1}$, with no downward component \citep{2014Sci...346A.315T}.
Spectroscopic observations from IRIS reveal that many of these jets are heated to $\sim10^5$ K \citep{2014Sci...346A.315T}. 
\cite{2016ApJ...820..124H} found a statistically significant match between automatically detected heating signatures in the corona, 
as observed in the AIA passbands, and quiet-Sun REs, with a minimum of 6\% of the detections at $\sim10^6$ K (AIA Fe IX 171) being attributable to REs. 

Despite a wealth of observations, the heating mechanism associated with chromospheric jets remains a mystery. 
Recent theoretical studies suggest that 
the Kelvin-Helmholtz Instability (KHI) could be a viable mechanism for the observed fast heating of chromospheric jets
\citep{2011AIPC.1356..106Z, 2015ApJ...802...26K,2015ApJ...813..123Z}. Mass flows in the chromospheric fine structures 
can create velocity discontinuities between the surface of the jets and surrounding media, which may trigger the KHI
in some circumstances depending on the directions of flows and magnetic fields.

\begin{figure}[t]
\begin{center}
\includegraphics[width=8.0cm]{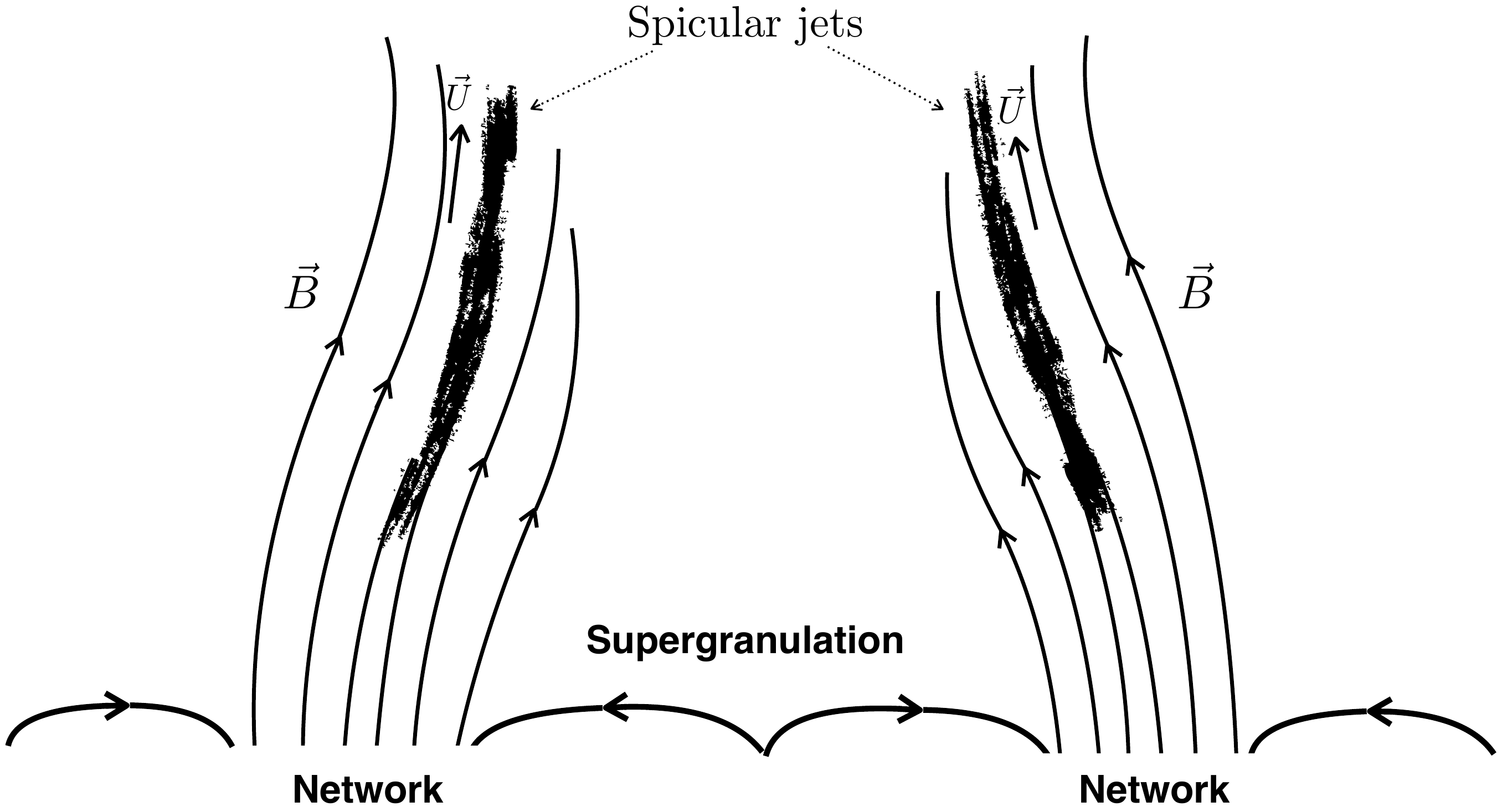}  %11
\end{center}
\caption{Simple schematic diagram of a  spicular jet considered as cylindrical, high-density, twisted magnetic flux tubes 
moving in an axial direction with respect to a twisted external field.}
\label{fig1}
%\label{fig5}
\end{figure}
 
The theory of KHI in solar atmospheric events has been intensively developed in recent years. 
It has been studied in spicular-like chromospheric jets \citep{2011AIPC.1356..106Z,2012A&A...537A.124Z}, magnetic tubes with partially ionized plasmas \citep{2012ApJ...749..163S,2015A&A...578A.104M}
in photospheric tubes \citep{2012A&A...547A..14Z}, and twisted and rotating magnetic jets \citep{2010A&A...516A..84Z,2014A&A...561A..62Z,2015ApJ...813..123Z}.
It has been shown that the energy conversion mechanism called resonant absorption \citep{2002A&A...394L..39G,2006RSPTA.364..433G,2011SSRv..158..289G}
may play an important role on the onset of KHI instability in magnetic flux tubes. 
I. e., the transverse MHD oscillations in coronal loops can lead to KHI 
through resonant absorption in narrow inhomogeneous layer that can deform the cross-sectional area of loops \citep{1994GeoRL..21.2259O,2008ApJ...687L.115T,2010ApJ...712..875S,2014ApJ...787L..22A}. 
Furthermore, it is suggested that the mode conversion from magneto-acoustic oscillations across the whole jet to small-scale localized Alfv\'en motions due to resonant absorption may also create the KHI through phase mixing \citep{1984A&A...131..283B}.
As well as the theory, recent observations show the presence of KHI in the solar corona, e.g., in prominences \citep{2010ApJ...716.1288B,2010SoPh..267...75R},  
coronal mass ejections and helmet streamers \citep{2011ApJ...734L..11O,2011ApJ...729L...8F,2013ApJ...767..170F,2013ApJ...766L..12M,2013ApJ...774..141F}.

\begin{figure}[t]
\begin{center}
\includegraphics[width=8.5 cm]{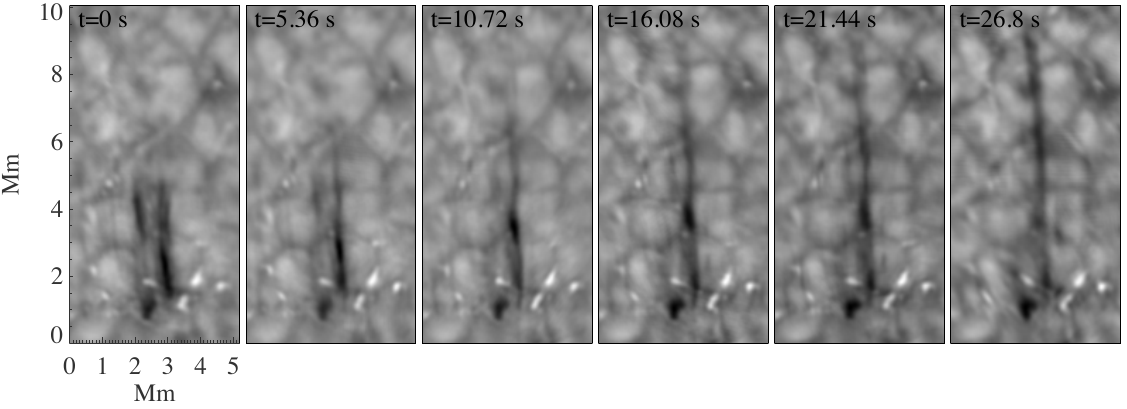} %11
\end{center}
\caption{Time-sequence of a typical jet in the red wing of $\mathrm{H\alpha+0.906 {\AA}~(41~km s^{-1})}$. 
The jet starts from the network brightening and moves upward with an apparent propagation speed of $\mathrm{\sim150~km~s^{-1}}.$}
\label{fig2}
%\label{fig5}
\end{figure}

Recently, \cite{2015ApJ...802...26K} estimated the growth rate of KHI in a transversely-moving REs using  
a simple slab model. They showed that the REs moving in the transverse direction 
could be unstable due to the KHI with a very short ($\mathrm{\sim5~s}$) instability growth time.  

In this paper we analyze the data presented in \cite{2015ApJ...802...26K} to study the dynamics of H$\alpha$ jets. 
We adopt the theoretical model of KHI for twisted magnetic jets in the solar wind 
developed by \cite{2014A&A...561A..62Z}, and derive a governing dispersion equation to investigate the stability of chromospheric H$\alpha$ jets.   
Using the theoretical model and observed jet parameters 
we estimate the growth time of the KHI in H$\alpha$ jets.
Furthermore, various energy dissipation mechanisms that could be responsible for the heating of chromospheric jets due to KHI
are investigated using the energy equation of the partially ionized plasma.
Observational evidence that could be considered as a signature of the KHI in the analyzed structures is also presented and discussed.

%%%%%%%%%%%%%%%%%%%%%%%%%%%%%%%%%%%%%%%%%%%%%%%%%%%%%%%%%%%%%%%%%%%%%%%%%%%%%%%%%%%%%%%%%%%%%%%%%%%%%%%%

\section{Kelvin-Helmholtz Instability: Theory}
\label{theo}
 
 \begin{figure*}[t]
\begin{center}
\includegraphics[width=16.86 cm]{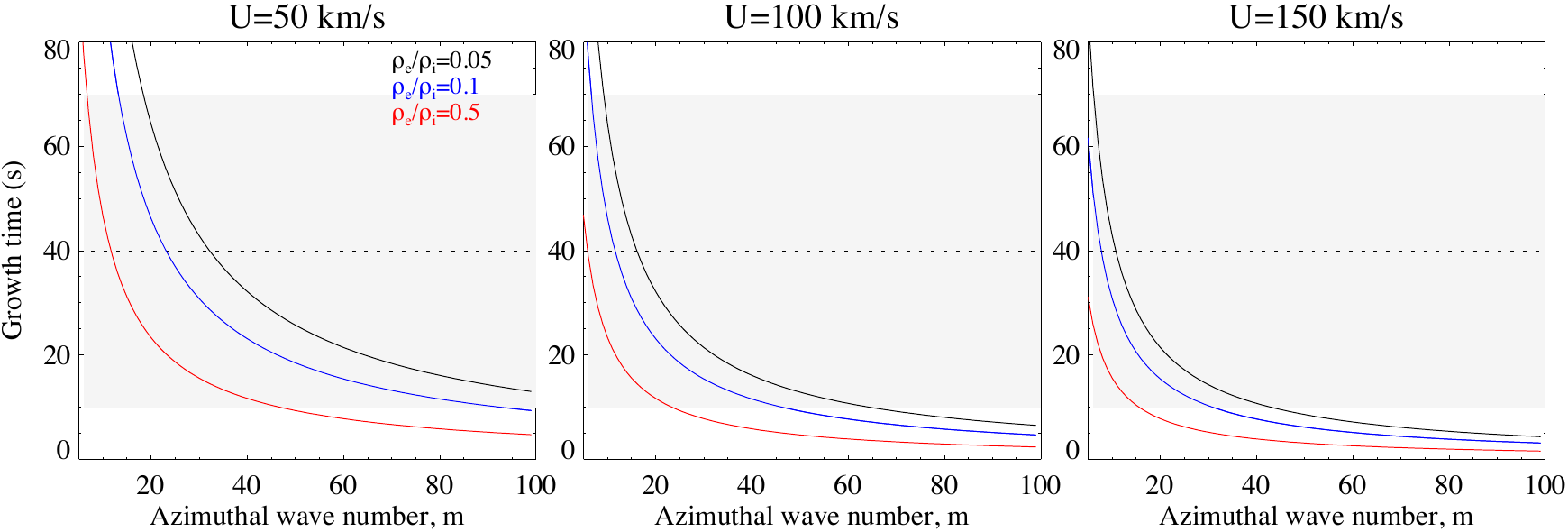}
\end{center}
\caption{KHI growth times as a function of $m$ for the twisted jets with different density ratios, $\rho_e/\rho_e,$ and flow speed $U$.
The growth times are derived from Equation (5) which is the solution of the dispersion equation for the high order of $m$.
The ratio of azimuthal and tangential components of magnetic field at the tube boundary
is $\mathrm{\xi=0.1}$. The grey-shaded areas indicate the typical range of lifetime ($\mathrm{\sim10 - 70~s}$) with the horizontal dotted lines indicate the average lifetime of REs.}
\label{fig3}
\end{figure*}

The KHI theory of twisted magnetized jets 
in the solar wind has been developed by \cite{2014A&A...561A..62Z}. 
They showed that, in the linear incompressible MHD limit,
KHI is suppressed in the twisted tube by the external axial magnetic field for sub-Alfv\'enic motions. 
However, even a small twist in the external magnetic field allows the KHI to be developed for any sub-Alfv\'enic motion.
Based on their approach, in our theoretical model the chromospheric jets are considered as cylindrical, high-density, twisted magnetic flux tubes 
with radius $a$ embedded in a twisted external field (Figure~\ref{fig1}).
We use a cylindrical coordinate system $\mathrm{(r, \phi, z)}$ and assume that the magnetic 
field has the following form: $\mathrm{B = (0, B_{\phi}(r), B_{z}(r))}$. The unperturbed magnetic field and pressure $(p_0)$ satisfy the pressure balance condition
$$
%\begin{equation}
\frac{d}{dr}\left(p_0+\frac{B^2_{\phi}+B^2_{z}}{8\pi}\right)=\left(\frac{B^2_{\phi}}{4\pi r}\right),
\label{eq6}
%\end{equation}
$$
The magnetic field inside and outside the jets is considered to be weakly twisted, thus not leading to the kink instability, 
and the tube moves along the axial direction with regards to the surrounding medium with speed $U$ (Figure~\ref{fig1}).   
In the cylindrical coordinate system the magnetic field inside and outside the tube 
is ${\bf B_i}=(0, Ar,B_{iz})$ and ${\bf B_{e}}=(0,Aa^2/r,B_{ez}(a/r)^{2})$, respectively, where $A$ is a constant.     
In this configuration internal and external magnetic fields are in the same direction and the twist of an 
internal and external fields at the tube/medium boundary are the same as well (Figure~\ref{fig1}). 
The unperturbed plasma pressure clearly depends on $r$ in order
to hold the pressure balance. We use linear ideal incompressible MHD equations, which consist of three components of the momentum equation, 
three of the induction equation and the continuity equation. The incompressible approximation considers negligible density perturbations and infinite sound speed. Therefore, the gradient of the pressure perturbation is retained in the momentum equation, while the time derivative of density 
perturbations is neglected in the continuity equation \citep{1961hhs..book.....C}. After straightforward calculations, the total (thermal + magnetic) pressure 
perturbations are governed by the Bessel equation \citep{1992SoPh..138..233G,2014A&A...561A..62Z,2015ApJ...813..123Z}%(Goossens et al. 1992, Zaqarashvili et al., 2014, 2015), 
which has a solution in terms of Bessel functions.
The continuity of the Lagrangian total pressure and displacement results in the dispersion equation
$$
\frac{\left(\left[\omega-k_{z}U\right]^{2}-\omega_{Ai}^{2}\right)F_{_{m}}(m_{i}a)-\frac{2mA\omega_{Ai}}{\sqrt{4\pi\rho_{i}}}}{\rho_{i}\left(\left[\omega-k_{z}U\right]^{2}-\omega_{Ai}^{2}\right)^{2}-\frac{4A^{2}\omega_{Ai}^{2}}{4\pi}}=
$$
\begin{equation}
=\frac{a^{2}\left(\omega^{2}-\omega_{Ae}^{2}\right)Q_{\nu}(m_{e}a)-2a^{2}\left(\omega^{2}-\omega_{Ae}^{2}\right)+\frac{2ma^2A\omega_{Ae}}{\sqrt{4\pi\rho_{e}}}}{a^{2}\rho_{e}\left(\omega^{2}-\omega_{Ae}^{2}\right)^{2}-\frac{4a^2A^2\omega^{2}}{4\pi}},
\label{eq1}
\end{equation}
where $\omega$ is the angular frequency, $\omega_{Ai}$ and $\omega_{Ae}$ are the Alfv\'en frequencies inside and outside the tube respectively,  
$m$ is the azimuthal wave number, $k_z$ is the longitudinal wavenumber, and $\rho_i$ and $\rho_e$ are the densities inside and outside the tube, respectively. 
Furthermore, 
$$
F_{m}(k_{i}a)=\frac{k_{i}aI_{m}'(k_{i}a)}{I_{m}(k_{i}a)},~Q_{\nu}(k_{e}a)=\frac{k_{e}aK_{\nu}'(k_{e}a)}{K_{\nu}(k_{e}a)}
$$ 
where $I_m$ and $K_{\nu}$ 
are modified Bessel functions of order $m$ and $\nu$, respectively, and 
$$
k^2_i=k^2_z\left[1-\frac{4A^2\omega^2_{Ai}}{4\pi\rho_i\left[(\omega-k_zU)^2-\omega^2_{Ai}\right]^2}\right],
$$
$$
k^2_e=k^2_z\left[1-\frac{4B_{e\phi}^2\omega^2}{4\pi\rho_i\left[\omega^2-\omega^2_{Ai}\right]^2a^2}\right]
$$
\citep[see][]{2014A&A...561A..62Z}.

The dispersion relation (1) is a transcendental equation
for the complex wave frequency, $\omega$, those imaginary part 
indicates an instability process in the system, in particular, the growth rate of unstable harmonics.
In the presented system, plasma flow in the twisted magnetic field can create velocity discontinuities between
the surface of the flux tube and surrounding media, which may trigger the KHI.
To simplify Equation (1) we  
consider perturbations with the wave vector nearly perpendicular to the magnetic field, i. e., 
${\bf k\cdot B_e=k\cdot B_i}\approx0$ which are vortices in the incompressible limit and have the strongest growth rate.
In the cylindrical coordinate system, this condition is expressed as
\begin{equation}
\mid m \mid \approx\frac{k_za}{\xi},
\label{eq22}
\end{equation}
where $\xi=aA/B_{iz}$ is the ratio of azimuthal and tangential components of magnetic field inside the tube. 
Furthermore, we assume that the azimuthal component (twist) of the magnetic field at the boundary of the flux tube is small ($aA\ll1$).
These assumptions reduce Equation (1) to the form:  
$$
\rho_{i}Q_{\nu}(k_{e}a)\omega^{2}-2\rho_{i}k_{z}UQ_{\nu}(k_{e}a)\omega +\rho_{i}k_{z}^{2}U{}^{2}Q_{\nu}(k_{e}a)-
$$
\begin{equation}
-2\rho_{i}\omega^{2}+4\rho_{i}k_{z}U\omega-2\rho_{i}k_{z}^{2}U{}^{2}-\rho_{e}F_{m}(k_{i}a)\omega^{2}=0,
\label{eq2}
\end{equation}
where $\nu\approx\sqrt{4+m^2}$ \citep[see][,for details]{2014A&A...561A..62Z}. 
Equation (3) is solved analytically 
for large azimuthal wave numbers, $m$.

\begin{figure*}[t]
\begin{center}
\includegraphics[width=16.3cm]{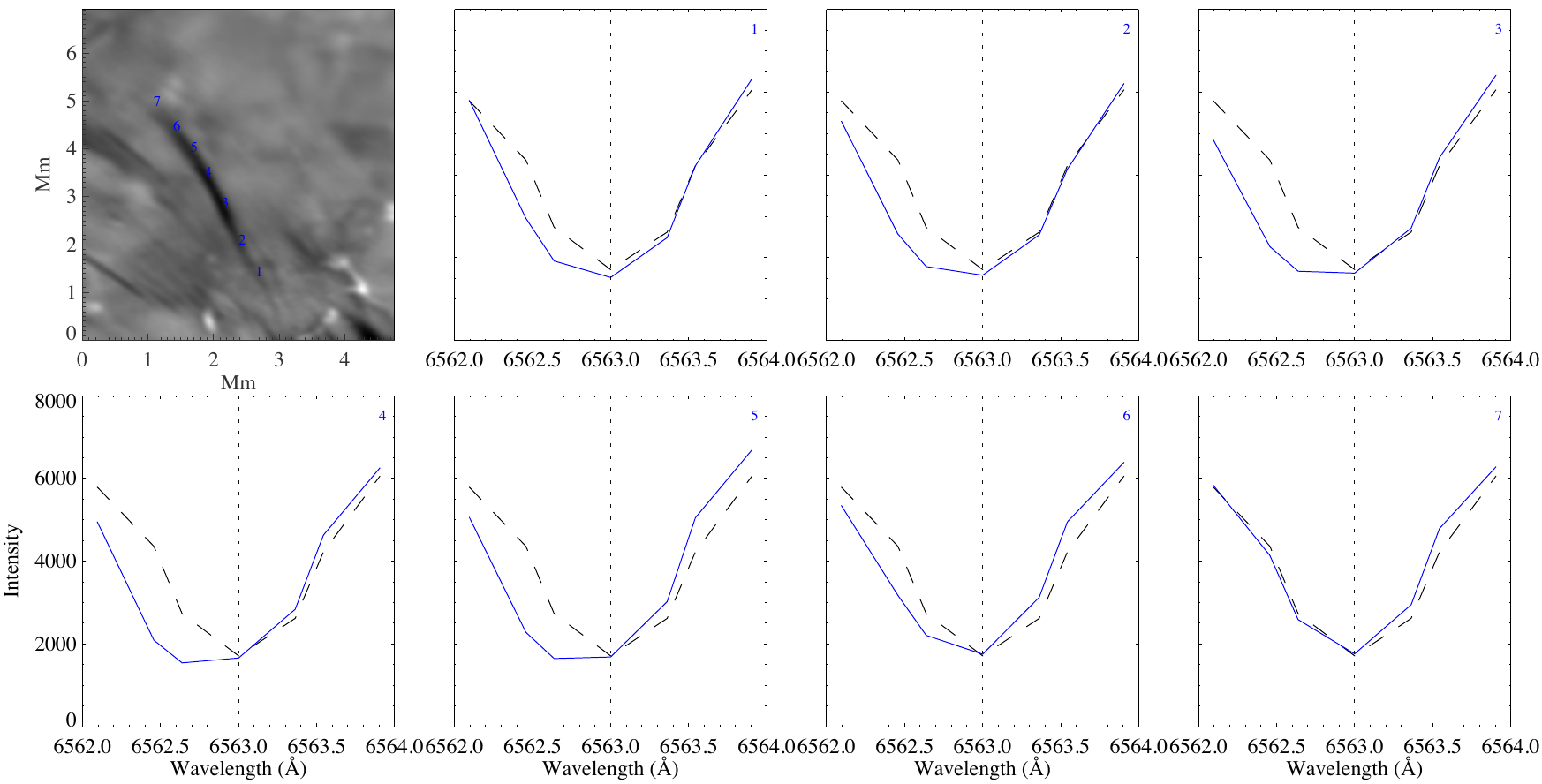}
\end{center}
\caption{H$\alpha$ line profiles (blue lines) of typical RBEs (detected in the $\mathrm{H\alpha+0.906 {\AA}}$) 
at different positions along its length. Numbers indicate the locations of line profiles. The black lines show the mean spectrum average over the field-of-view for reference.}
\label{fig4}
%\label{fig5}
\end{figure*}

For large azimuthal wave numbers, $m$ and $\nu$, asymptotic expansion of the modified Bessel functions for 
large order can be implemented in the dispersion equation \ref{eq2} \citep[see, e. g.,][]{1965hmfw.book.....A}.  
This gives $F_{m}(k_{i}a)\approx1$, $Q_{\nu}(k_{e}a)\approx-1$ and Equation (3) can be simplified to
\begin{equation}
\left[1+\frac{1}{3}\frac{\rho_{e}}{\rho_{i}}\right]\omega^{2}-2k_{z}U\omega+k^2_zU^{2}=0.
\label{eq5}
\end{equation}
The solution of Equation~\ref{eq5} is always complex number 
which indicates that the perturbations are unstable to KHI for large azimuthal wave numbers. 
The growth rate of the unstable harmonics is an imaginary part of the solution of Equation~\ref{eq5},
\begin{equation}
\omega_{i}=\frac{\sqrt{\rho_e/3\rho_i}}{1+\rho_e/3\rho_i}  
\frac{\mid m\mid \xi}{a}U.
\label{eq6}
\end{equation}

Equation~\ref{eq6} shows that the growth rate of KHI depends on the flow speed, 
radius of the tube, the density contrast, the ratio of azimuthal and tangential components of magnetic field, 
and the spatial scale of perturbations in the azimuthal direction ($m$). 
It can be used to estimate KHI growth times of specific jets in the solar chromosphere. 

We note that the dispersion equation similar to equation (4) for the twisted 
magnetic flux tube has been solved analytically by \cite{2014A&A...561A..62Z}.
They showed that the harmonics with small azimuthal wavenumbers have smaller growth rates compared to those with larger ones. 
Therefore, we use the large azimuthal wavenumber limit in the dispersion equation (3) in order to consider the unstable harmonics with larger growth rates. There are two additional reasons why the harmonics of large azimuthal wavenumbers are important.
First, numerical modeling shows that the photospheric/chromospheric magnetic flux tubes are 
developing KHI with large azimuthal wavenumbers $(m>5)$ \citep[see e.g.,][]{2008ApJ...687L.115T,2014ApJ...787L..22A,2015ApJ...809...72A,2016MNRAS.459.2566M}. Second, a large azimuthal wavenumber obviously yields smaller wavelength, which is important for plasma heating (see subsection 4.2).

We note that the incompressibility limit we have used in this model is a valid approximation as compressibility does not have a strong influence on the dynamics of KHI 
for the harmonics considered here.
\cite{2012ApJ...749..163S} has shown that the effect of compressibility on KHI is controlled by 
the ratio of longitudinal and tangential components of wave vector ($k_y/k_z$) with respect the magnetic field. 
An equilibrium magnetic field is straight and directed along z in the configuration used by \cite{2012ApJ...749..163S}.
For large $k_y/k_z$, the growth rates of KHI for compressible and incompressible cases are almost similar \citep[see Figure 1 of][]{2012ApJ...749..163S}.  
For $k_y/k_z\rightarrow\infty$, which are harmonics propagating perpendicular to the magnetic field, both compressible and incompressible results agree. 
Therefore, compressibility plays no role when the wave vector is perpendicular to the magnetic field/flow direction. 
\cite{1979ApJ...227..319W} has also noted that the perturbations with $\mathrm{k_z/ k_y \ll1}$, where magnetic field is directed along z, involve negligible compression.
In our model we have considered harmonics 
with wave vector perpendicular to the magnetic field (pure vortices) and hence, incompressibility is a valid approximations. 

Furthermore, our MHD model is developed in the single fluid approach, where ion-electron and neutral gasses are not considered as a separate fluids. 
However, for the time-scales that are longer than ion-neutral collision time the single fluid model is valid as collisions between 
neutrals and charged particles lead to the rapid coupling of the two fluids \citep[see details in][]{2011A&A...529A..82Z}.   
\cite{2015A&A...573A..79S} has shown that in the upper chromosphere where 
ion-neutral collision frequency could be smaller than the ion-ion and ion-electron collision frequencies,
the two-fluid model is better approximation.
Soler at al 2015 has shown that the two-fluid model is valid whenever ion-neutral collision frequency is smaller than the ion-ion and ion-electron collision frequencies, which could be the case in the upper chromosphere.  
However, for spicular jets ion-neutral collision frequancy is $\mathrm{10^3~Hz}$. Corresponding spatial scales are around 50 m 
for the chromospheric Alfv\'en speed ($\mathrm{\sim50~km~s^{-1}}$).  While we deal with spatial scales of $\mathrm{>50~km}$ single fluid approach is a valid approximations.

%%%%%%%%%%%%%%%%%%%%%%%%%%%%%%%%%%%%%%%%%%%%%%%%%%%%%%%%%%%%%%%%%%%%%%%%%%%%%%%%%%%%%%%%%%%%%%%%%%%%%%%%

\section{Observations and data reduction}
\label{sect:setup}

The data presented here have been partly studied in \cite{2015ApJ...802...26K}. 
Observations were undertaken between 09:06 and 09:35 UT on 2013 May 3 at disk centre with the CRisp Imaging
SpectroPolarimeter \citep[CRISP;][]{2006A&A...447.1111S,2008ApJ...689L..69S} instrument, mounted on the Swedish 1-m Solar Telescope \citep[SST;][]{2003SPIE.4853..341S}
on La Palma.  Adaptive optics were used throughout the observations, consisting of a tip-tilt mirror 
and a 85-electrode deformable mirror setup that is an upgrade of the system described in \cite{2003SPIE.4853..370S}. 
The observations comprised of spectral imaging in the H$\alpha$~6563~{\AA}, and Fe {\sc{i}} $\mathrm{6302~{\AA}}$ lines.
All data were reconstructed with Multi-Object Multi-Frame Blind Deconvolution \citep[MOMFBD;][]{2005SoPh..228..191V}. 
We applied the CRISP data reduction pipeline as described in \cite{2015A&A...573A..40D}, including small-scale seeing compensation as in \cite{2012A&A...548A.114H}.
Our spatial sampling is 0$''$.0592 pixel$^{-1}$
and the spatial resolution approaches the diffraction limit of the telescope at this wavelength ($\sim 0 " 16$) for a large and steady portion of the images throughout the time series.
The H$\alpha$ line scan consist of 7 positions (-0.906, -0.543, -0.362, 0.000, 0.362, 0.543, +0.906 {\AA} from line core), corresponding to a range of -41 to +41 km s$^{-1}$ in velocity.
A full spectral scan had an acquisition time of 1.3 s, which is also the temporal cadence of the H$\alpha$ time series.
We made use of CRISPEX \citep{2012ApJ...750...22V}, a versatile widget-based tool for effective viewing and 
exploration of multi-dimensional imaging spectroscopy data.

%%%%%%%%%%%%%%%%%%%%%%%%%%%%%%%%%%%%%%%%%%%%%%%%%%%%%%%%%%%%%%%%%%%%%%%%%%%%%%%%%%%%%%%%%%%%%%%%%%%%%%%%

\section{Analysis and Results}

\cite{2015ApJ...802...26K} studied the dynamics of REs in the dataset presented here. 
They detected a total of 70 RREs in the far red wing at H$\alpha$ + 0.906 {\AA} and 58 RBEs in the far blue wing at H$\alpha$  - 0.906 {\AA},
and analyzed their lifetime, length, width, 
speed of their apparent motion, line-of-sight (LOS) velocity, and transverse velocity for each individual detection.  
The statistical study of their properties showed that the lifetimes of the RBEs/RREs range from 10 s  to 70 s, with a median of 40 s, the lengths are in range $\mathrm{2 -9}$ Mm with a median $\sim$3~Mm and the widths 
between the $\mathrm{150-500~km}$, with a median of $\sim$260~km \citep[see Figure 2 and 4 in ][]{2015ApJ...802...26K}.

Detected structures also display non-periodic, transverse motions perpendicular to their axes at speeds of $\mathrm{4-31~km~s^{-1}}$.
\cite{2015ApJ...802...26K} proposed that the transverse motions of chromospheric flux tubes can develop the KHI at the tube boundaries due to the velocity discontinuities between the surface 
of the flux tube and surrounding media. 
Using  a simple slab model they showed that the REs moving in the transverse direction with speed comparable to the local Alfv\'en speed 
could be unstable due to the KHI with a very short 
instability growth time (see \cite{2015ApJ...802...26K} for further details).

Many of the detected structures appear as high-speed jets that are directed outwardly from a magnetic bright point.  
Figure~\ref{fig2} shows the temporal evolution of a typical jet detected in the red wing at H$\alpha$ + 0.906 {\AA}.
The jet starts near the bright point and moves upward with an apparent propagation speed of $\sim$150  km s$^{-1}$. 
Apparent velocities of the REs projected on the image plane are in the range 50 - 150  km s$^{-1}$ (see Figure 4 in \cite{2015ApJ...802...26K}) 
which are Alfv\'enic and super-Alfv\'enic in the chromosphere.

Some recent observations show that  REs and type II spicules display torsional/twisting motions during their evolution \citep{2012ApJ...752L..12D,2014Sci...346D.315D}.  
If we consider the chromospheric jets as a cylindrical, twisted magnetic flux tubes 
moving along their axis with respect to the external twisted field, then  
(as we show in Section~\ref{theo}) these structures may be unstable to KHI.

The  growth time, $T$, of the KHI is given by  $T=2\pi/\omega_i$, 
where $\omega_i$ is defined by Equation~(5).
If the KHI is a viable mechanism for the heating/disappearance of REs then the growth time for the instability should be comparable to or smaller than the structure's lifetime.   
To estimate the dependence of the growth time of KHI on the azimuthal wave number, 
we employ three different values of the measured apparent speeds, $\mathrm{U=50, 100, and~150~km~s^{-1}}$, and the median value of radius of REs, $\mathrm{a=130~km}$, in Equation~(5). 
The results are presented in Figure~\ref{fig3},
which represents the growth times derived from Equation~(5) which is the solution of the dispersion equation for the high order of $m$.  
Unfortunately, the present observations do not allow a determination of the density ratios between the outside and the inside the jet. 
However, as spicular jets are observed as overdense features 
\citep[e. g.,][]{1968SoPh....3..367B}, values of growth time are computed for three different density ratios  
$\mathrm{\rho_e/\rho_i=0.1, 0.5,~and~0.05}$. 
For the ratio of azimuthal and tangential components of magnetic field at the tube boundary in Equation~(5),  
we took $\mathrm{\xi=0.1}$ which implies that the magnetic field is only slightly twisted.
We see that the 
the instability growth times strongly depend on the flow speed and density ratio.  
Higher flow speeds produce faster growth, and higher density ratios slow the growth of the instability (Figure~\ref{fig3}).
The flow speed, $\mathrm{U\sim150~km~s^{-1}}$,
for the harmonics with azimuthal mode numbers $5\leqslant m\leqslant 40$ leads to growth times of $T\sim5 - 85~s$, comparable to the lifetime of REs ($\sim$10-100~s). 
For speeds in the range $\mathrm{U\sim50-100~km~s^{-1}}$, the KHI growth times for the same harmonics are longer ($T\sim45 - 250~s$)
The higher harmonics have very short instability growth times. 
For harmonics with $m\geqslant50$ the growth times are  $\mathrm{1-5~s}$ depending on the flow speed and density ratio.    

\begin{figure*}[t]
\begin{center}
\includegraphics[width=15.3cm]{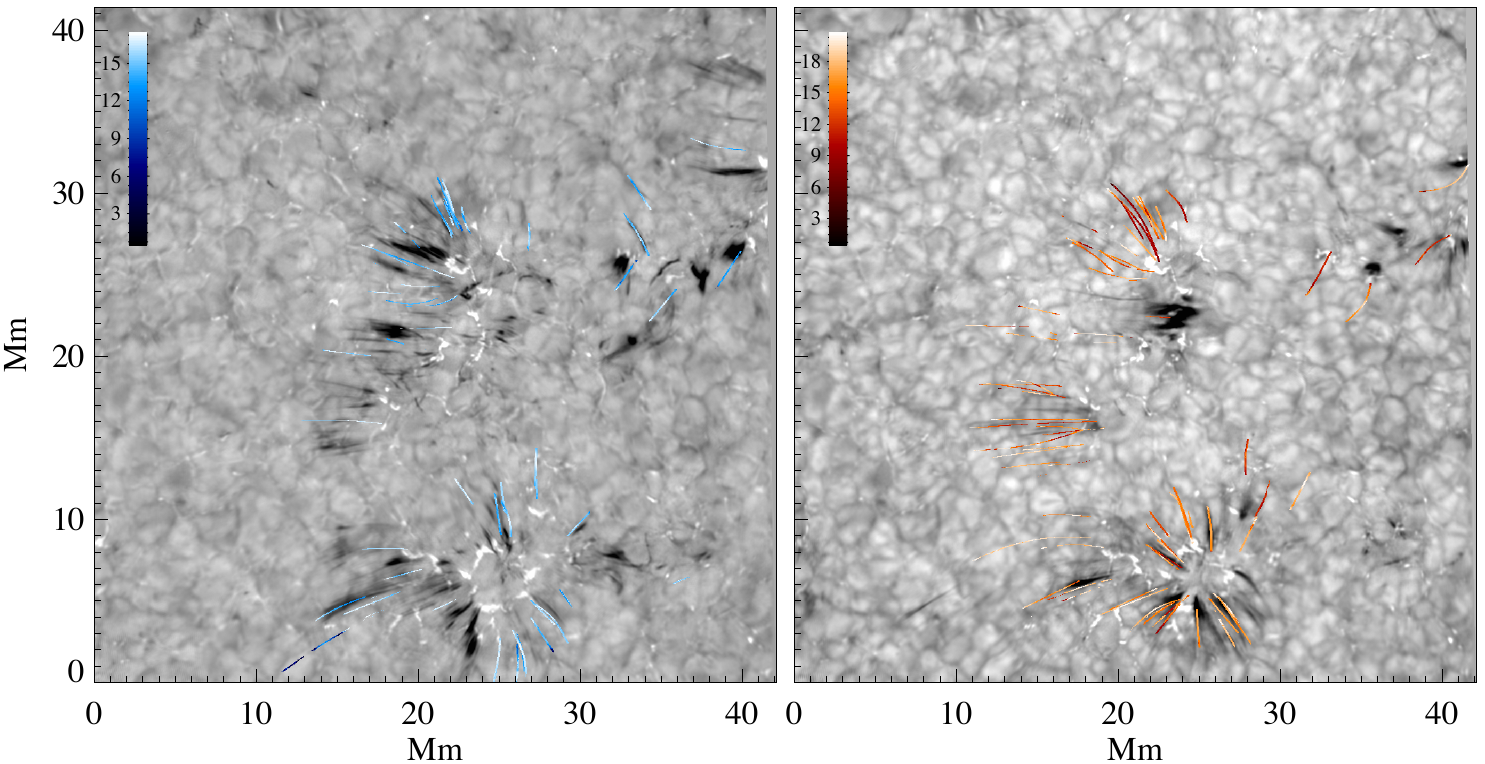}
\end{center}
\caption{H$\alpha$ Doppler widths along RBEs (left panel) and RREs (right panel) over-plotted at their locations in the field-of-view, with the color scale indicating the velocity in $\mathrm{km~s^{-1}}$.}
\label{fig5}
\end{figure*}

The generated KH vortices may lead to enhanced MHD turbulence near the boundaries of REs,
which could be observed as non-thermal broadening in spectral line profiles.    
We therefore study the evolution of the H$\alpha$ line profiles for the detected REs. 
Figure~\ref{fig4} represents the line profiles of typical RBEs at different positions along its length (blue lines) 
together with a mean spectrum average over the field-of-view for reference (black lines).
We see that the line profiles have an extended absorption wing on the blue side of the core, and they are wider all along the structures' length (Figure~\ref{fig4}).
The line profile at position 7, which is from outside of the structure, is very close to the mean profile. 
Values of Doppler width for each  individual feature are calculated from Doppler signals provided by their H$\alpha$ spectral profile
using the second moment with respect to wavelength method described by \cite{2009ApJ...705..272R}.
The Doppler width of the structures is computed for every pixel and every frame of the REs. Full paths of the REs were selected manually, 
but the computations are only performed where the wing shows an opacity dip according to the criteria described in \cite{2009ApJ...705..272R}.
The computed Doppler width for each individual detection is shown in Figure~\ref{fig5}, where the final values for each pixel detected as valid, are used. 
Figure~\ref{fig5} indicates that the Doppler widths are increased for the REs,  with average width for RBEs and RREs found to be $\mathrm{\sim10~km~s^{-1}}$ and $\mathrm{\sim12~km~s^{-1}}$, respectively. 
It should be noted that these estimates do not have a large accuracy as 
the method employs the line spectrum to separate and compute the LOS Doppler velocity and Doppler width components using first and second moments of the profile. 
Another limitation of the method is that it computes the line width for only the blue or red component of the line wing, as normally there is not any spectral signature of the detection at the opposite side of the wing. 
Nonetheless, the analysis shows clearly that the REs detected in red- and blue-wings in H$\alpha$ have enhanced line widths all along the structure's lengths (Figure~\ref{fig5}).  
This suggest that, if line broadening is produced by turbulent motions associated with KHI vortices, 
then the longitudinal scale of the KHI should be much shorter than the length of the structures.

\begin{figure}[t]
\begin{center}
\includegraphics[width=8.5cm]{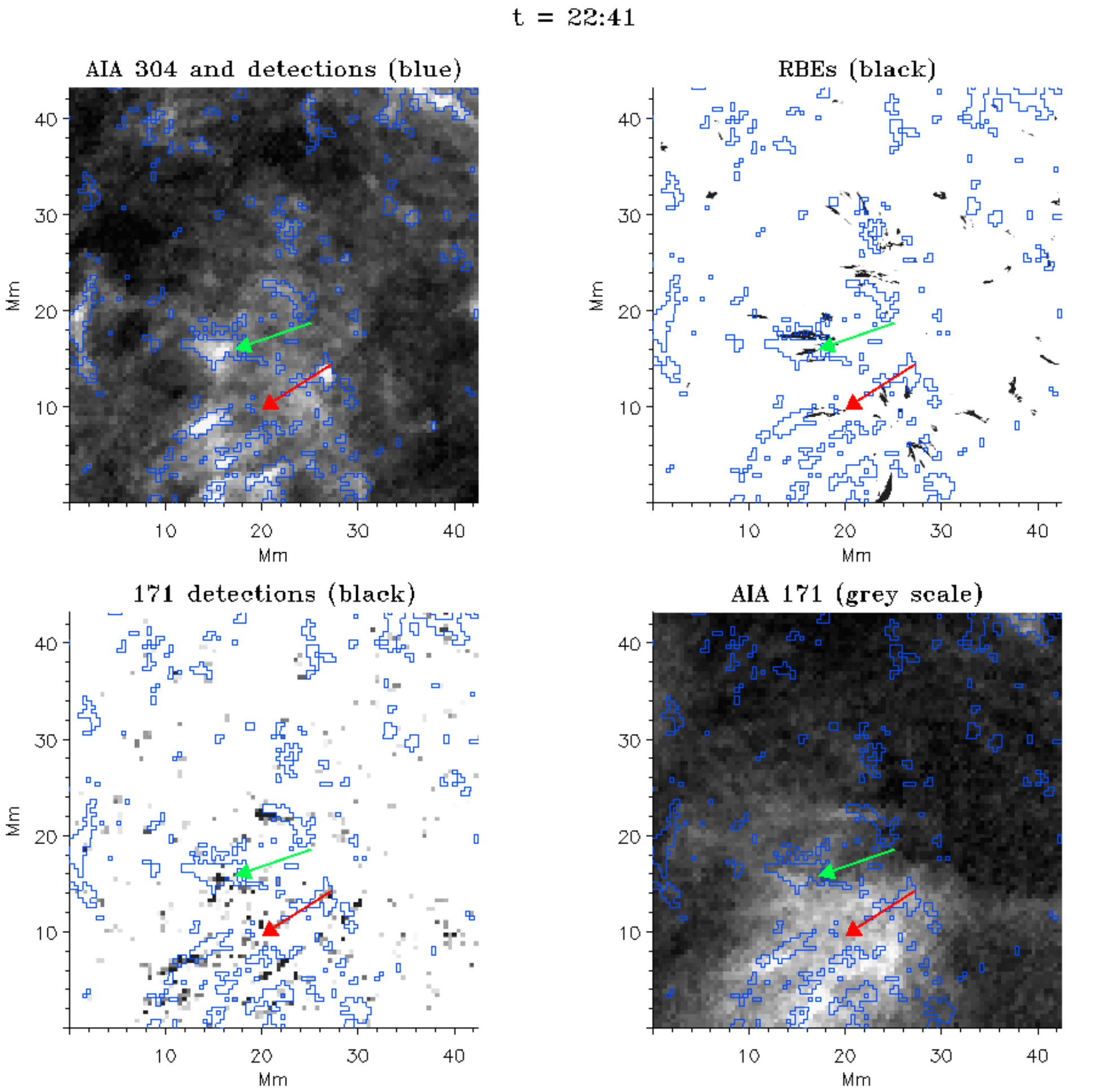}
\end{center}
\caption{Panels extracted from the first time-series of \cite{2016ApJ...820..124H} where we highlight two examples of jets crossing the chromosphere and the corona (one colour per jet). 
The arrows point at the same point across all channels: the far-right signature of each jet which is visible as a tip from high-contrast RBEs (see panel labelling). 
The thin dark path traced by the red-arrow RBE is seen connecting with an elongated blue contour. Such blue contour comes from an automated detection 
algorithm sensitive to 50 s variations in the AIA 304 \AA\ channel. However, for this frame, such variation is also visible as a bright patch in the AIA 304 \AA\ intensity (top left panel). 
Such a contour connects, at the other end of the elongation, with a similarly detected 171 \AA\ channel signature (which is also visible in the 171 \AA\ intensity). 
This indicates a phased progression across all channels, which most certainly involves changes in 
temperature and likely ionization state, considering the formation ranges involved. The green arrow jet follows the same pattern, with the difference that all 
signatures are nearly overlapped as their shape is much narrower across all channels (indicating a more vertical propagation),
and the 304 \AA\ detection contour encompasses the RBE, the 171 detection, the 171 brightening, and another co-propagating RBE.}
\label{fig55}
%\label{fig5}
\end{figure}

The wavelength of the KHI perturbation can be determined using $\lambda=2\pi/k_z$, where $k_z\approx| m | \xi/a$ is defined from Equation~(2).
It shows that the harmonics with high order of azimuthal wavenumbers, which have faster growth time,  
correspond to KHI perturbations with short wavelengths.  
For $\mathrm{a=130}$, $\xi=0.1$ and $30\leqslant|m|\leqslant100$, the wavelength is in the range $\mathrm{\lambda\simeq70-270~km}$, much less than the RE measured lengths ($\mathrm{\sim3.5~Mm}$).

It should be also noted that the line broadening can be caused by rotation/torsional motion of the structures around the axis. 
This may happen when the spatial resolution of observations is less than the width of structures, and one cannot resolve the opposite directed 
motions at the left and right sides of structures.  
The spatial resolution of the presented H$\alpha$ observations is around half  ($\mathrm{\sim120~km}$) 
of the measured average width ($\mathrm{\sim260~km}$) of REs, and hence the rotational motion should be resolved. 
However, we have not detected this in the current dataset, which suggests that the MHD turbulence/heating is the ultimate source of line broadening.

As well as small-scale REs, larger-scale jets have also been observed near the center of the rosettes in the blue wing of H$\alpha$, 
and an example is shown in Figure~\ref{fig6}. It appears 
in the far blue wings at $\mathrm{ - 0.906~{\AA}~and~ - 543~{\AA}~from~the~H\alpha}$ line core 
with speed of $\sim$34 km s$^{-1}$ and width $\sim$800 km.  
The structure does not have a horizontal extension, indicating that the jet was ejected vertically to the solar surface along the LOS (Figure~\ref{fig6}). 
At the start of the ejection the jet has a circular shaped top. However, 
the structure rapidly develops small vortex-shaped features 
on time scales of tenth of a second near its boundary (Figure~\ref{fig6}).   
These features may correspond to the projected vortex-flows of the KHI perturbations.
Unfortunately, $\sim$75~s after the jet's first appearance, the observation stopped and we are unable to study the full lifecycle of the jet.  
However, in the last images the jet has almost disappeared from the far blue wing image at $\mathrm{H\alpha-0.906~{\AA}}$ (Figure~\ref{fig6}),
indicating that it is in the disappearing phase of its life. 
The growth time of the KHI as a function of azimuthal wave number, $m$, obtained from Equation~(7)
for the jet with radius, $\mathrm{a=400~km}$,
upflow speed $\mathrm{U=34~km~s^{-1}}$, and the ratio of azimuthal and tangential components of magnetic field $\xi=0.5$, is presented in Figure~\ref{fig7}. 
It could be estimated that at around $\mathrm{t=37.52~s}$, the erupting structure has around $\mathrm{~8}$ resolved projected vortex-like features/azimuthal nodes around its boundary (see Figure~\ref{fig6}). 
For the harmonic $m=8$ the growth time $T\sim52 - 145~s$, depending on the values of the density ratio outside and inside the tube (Figure~\ref{fig7}).  
This growth time appears to be consistent with the time scale ($\sim$37~second) taken for the structure to develop vortex like features (Figure~\ref{fig6}). 
$\xi=0.5$ is well below the threshold for kink instability, which starts around $\xi=2$, therefore the dispersion equation is still correct for such a twist.
However, for lower $\xi$, i. e.,  $\xi=0.1$  the growth time is much longer,  $T\sim260 - 700~s$, depending on the values of the density ratio.

%%%%%%%%%%%%%%%%%%%%%%%%%%%%%%%%%%%%%%%%%%%%%%%%%%%%%%%%%%%%%%%%%%%%%%%%%%%%%%%%%%%%%%%%%%%%%%%%%%%%%%%%

\subsection{Heating signatures in the transition region and corona}

Recent observations have revealed that chromospheric type II spicules are rapidly heated to transition region and even coronal temperatures. 
Using the dataset presented here, \cite{2016ApJ...820..124H} showed the connection between REs and coronal structures. 
The automated detections of coronal transients was achieved by employing a running difference technique, aimed 
to have the highest sensitivity for 
50 s transients. By using a robust statistical analysis based on Chernoff bounds, 
 absolute minimum values of 6\%\ of the SDO/AIA 171 \AA\ channel events and 11\%\ of the 304 \AA\ events were shown to have an RE counterpart.  
The probability of the observed matches being due to noise or random chance was demonstrated 
to be lower than $10^{-40}$. 
Figure~\ref{fig55} shows some specific examples, where the structures are clearly seen in all channels 
including the RE maps, 304 \AA\ and 171 \AA\ lines \cite[for more example see Figure~3 to 9 of][]{2016ApJ...820..124H}.
Such particular matches were also found by \cite{2014ApJ...792L..15P} between 304 \AA\ and type II spicules in the $H_{\alpha}$ line,
 and by \cite{2015ApJ...799L...3R} in transition region channels.

%%%%%%%%%%%%%%%%%%%%%%%%%%%%%%%%%%%%%%%%%%%%%%%%%%%%%%%%%%%%%%%%%%%%%%%%%%%%%%%%%%%%%%%%%%%%%%%%%%%%%%%%

\subsection{Heating of chromospheric jets due to KHI}

\begin{figure*}[t]
\begin{center}
\includegraphics[width=16.5cm]{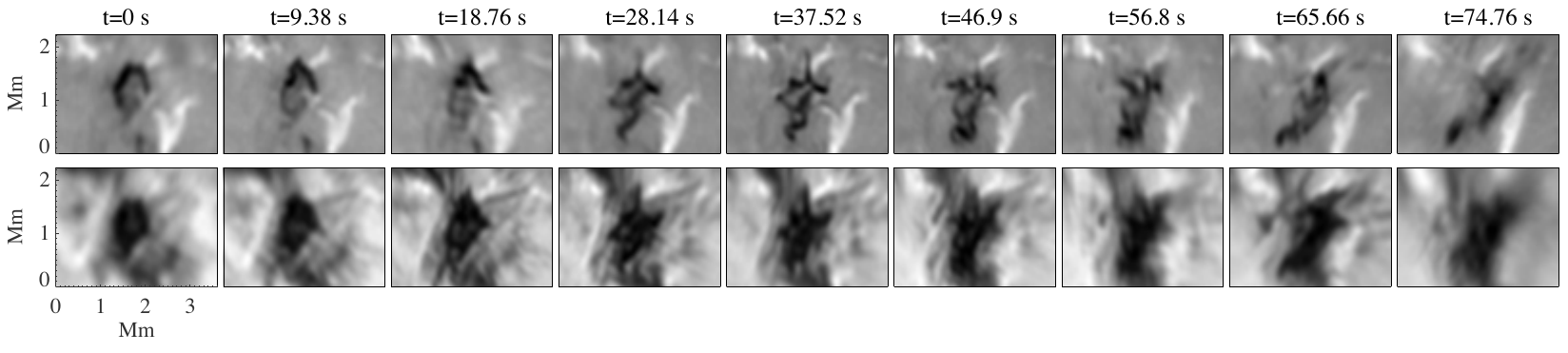}
\end{center}
\caption{Sequence of frames showing the temporal evolution of the jet ejected along the LOS
in $\mathrm{H\alpha-0.906~{\AA} (top) ~and~H\alpha - 0.543~{\AA}~of~H\alpha}$ (bottom) images. Initially the jet has circular top, however, it 
rapidly develops azimuthal nodes around its boundary.}
\label{fig6}
%\label{fig5}
\end{figure*}

The rapid heating of chromospheric jets is an unsolved problem. An initial explanation is that KHI destroys the jets by the rapid mixture of chromospheric and coronal plasmas, 
which could lead to the rapid disappearance of the jets. However, in this case the jets must be destroyed completely (as indicated
by numerical simulations) 
and should not appear in hotter spectral lines, in clear contrast with observations. 
On the other hand, large-scale KH vortices may transfer their energy into 
smaller scales through nonlinear cascade, where it could be transformed into heat. Furthermore, several dissipation mechanisms can produce a direct energy transfer from KH vortices into heat without non-linear cascade.

In our simplified ideal MHD model the various dissipation effects, such as 
diffusivity, viscosity, radiative cooling, and thermal conduction, are neglected.
These MHD processes
could have an important impact on the lifetime and stability of chromospheric structures. 
Unfortunately, the exact effects of these dissipation processes  
are not well studied in spicule-like jets.
However, the lifetimes of the small-scale REs are too short to be explained by these dissipation processes. 
\cite{2014ApJ...785..109L} estimated that 
only structures that have a radial scale less than 5 km are expected to fail the frozen field condition and diffuse within their lifetime ($\sim$100~s).
This suggests that the time taken for chromospheric jets with a typical width of $\sim$250~km to diffuse is
 expected to be much longer than their thermodynamical lifetime. 

As spicular jets are chromospheric features, thermal conduction probably has a small effect on their 
dynamics over such short timescales ($\sim$100~s). 
The heating/cooling timescale due to thermal conduction can be written as 
$$
\tau_c\approx{\frac{3n k_BL^2}{2/7\kappa T^{5/2}}}
$$
where $\mathrm{\kappa = 1.1 \times 10^{-6}~erg~s^{-1}~cm^{-1}~K^{-1}}$ is the Spitzer thermal conductivity, $k_B$ the Boltzmann constant,
$T$  the temperature, and $n$ and $L$ the number density and half-length of the structure, respectively \citep{1997ApJ...478..799C,2004A&A...424..279T}.   
For the typical number density of spicular jets, $\mathrm{n\sim10^{11}~cm^{-3}}$, half-length $\sim$~1.5 Mm and transition region temperature $\sim\mathrm{10^{5}~K}$, the
thermal conduction timescale is around 26 hours, 
much longer than the lifetime of the observed features.

\cite{2004A&A...424..279T} have shown that the energy flux due to radiative cooling in chromospheric H$\alpha$ mottles
is much higher than the conductive flux. 
Typical radiative cooling times are expected to be of the order of minutes \citep{1989ApJ...346.1010A,2006ASPC..354..259J}, 
 still longer than the dynamical timescales of REs.

It should be noted that in the theoretical model we have assumed a fully ionized plasma. However, 
the chromosphere is only partially ionized, 
with the ratio of neutral atoms to electrons continuously increasing from almost zero at the transition region to $\sim$10$^4$ at the photospheric base. 
The interaction of ions and neutral atoms influences basic plasma processes leading to the damping of MHD waves and/or electric currents 
\citep{2011JGRA..116.9104S,2014ApJ...796L..23S,2012ApJ...747...87K}. 
Both types of spicules, as well as RREs/RBEs, have to be composed of partially 
ionized plasmas, and therefore ion-neutral collisions may lead to the heating of KH vortices and consequently the structure itself. 
However, the detailed study of the heating process is not the main goal of the present paper, and the characteristic heating times of different dissipation mechanisms can be estimated through the energy equation. 
 
The equation of energy in a partially ionized plasma can be written as \citep[see e.g.,][]{2011A&A...534A..93Z}     
$$
{{\partial p}\over {\partial t}}+({\bf V}\cdot\nabla)p+\gamma p \nabla \cdot {\bf V}=
$$
\begin{equation}
(\gamma-1){{\alpha_{ei}}\over {e^2n^2_{e}}}{\bf j}^2+(\gamma-1)\alpha_{in}{\bf w}^2+(\gamma-1)Q_{visc}+Q_{cool}, 
\label{energy}
\end{equation}
where $p$ and $\bf V$ are the total pressure and the velocity of protons, electrons, and neutral hydrogen atoms respectively,  
$\alpha_{ei}$  and $\alpha_{in}$ 
the coefficients of friction between ions and electrons, and ions and neutral hydrogen atoms, respectively, 
$n_{e}$ the electron number density, ${\bf j}=(c/4\pi) \nabla \times {\bf B}$ the current, ${\bf w}=\bf V_i-\bf V_n\approx (\xi_n/c \alpha_{in}) {\bf j}{\times}{\bf B}$ 
 the velocity difference between protons and neutrals, $\xi_n$ the ratio of neutral to total particle density, $\gamma=5/3$ 
 the ratio of specific heats, and $c$ the speed of light. 
The quantity $Q_{visc}=-\pi_{\alpha \beta}W_{\alpha \beta}$ is the heating due to viscosity, where $\pi_{\alpha \beta}$ is the viscous stress tensor and $W_{\alpha \beta}$ the shear velocity tensor \citep{1965RvPP....1..205B}. %(Braginskii 1965). 
On the right-hand side of Equation 6, the first, second and third terms are associated with the heating by Joule, ion-neutral collision and viscosity, respectively. 
The last term, $Q_{cool}$, 
is associated with several cooling processes such as  thermal conduction and radiation. 
However, as discussed above, these processes are unimportant over the timescales of REs and hence may be ignored.
The ratio of the first term (associated with Joule heating) to the second (heating by ion-neutral collisions) 
on the right-hand side of Equation~6 is approximately  
\begin{equation}
{{Q_{ei}}\over {Q_{in}}}\approx {{m_i m_e c^2 \delta_{ei}\delta_{in}}\over {e^2 \xi_n B^2}}, 
\label{rat1}
\end{equation}
where $m_i$ and $m_e$ are the proton and electron masses, respectively, 
$\delta_{ei}$ and $\delta_{in}$ the electron-ion and ion-neutral collision frequencies, respectively,  
$e$ the electron charge, and $B$ the magnetic field strength. 

It must be noted that the ion-neutral collision frequency $\delta_{in}$ is generally different from that for neutral-ion collisions 
$\delta_{ni}$ in partially ionized plasmas.
They differ by the factor of ionization fraction. \cite{2011A&A...529A..82Z} suggested 
that the actual physical meaning of the collision time is expressed by the formula $1/(\delta_{in}+\delta_{ni})$, 
which shows the timescale over which the relative velocity between ions and neutrals (${\bf w}=\bf V_i-\bf V_n$) decreases exponentially. 
Hence, it shows the timescale of energy exchange between ions and neutrals. This suggestion was fully verified by recent numerical simulations \citep{2016ApJ...818..128O}. 
Unfortunately, observations do not allow us to estimate the precise 
ionization fraction in chromospheric jets.   
However, we know that the ionization fraction is changing from almost zero at the transition region ($T\sim10^5~K$) to $\sim$1 
at the photospheric base ($T\sim5\times10^3~K$). Based on some parameters given by the cloud model, 
\cite{1997A&A...324.1183T} derived an ionization degree for hydrogen in chromospheric jets (mottles) of 
around 0.65. \cite{2011ApJ...733L..15V} 
also estimated the ionization degree of hydrogen atoms along a spicule. 
They showed that the lower part of the spicule has a small ionization degree (0.01 -- 0.1), but the 
 upper half has a value close to 0.5. 
Therefore, in the following we adopt an ionization degree of $\sim$0.5, which yields the same values of ion-neutral and neutral-ion collision frequencies.
From \cite{2015A&A...573A..79S}, at a height of 2000 km and for $\xi_n$=0.5,
$\delta_{ei}$=10$^7$ Hz and $\delta_{in}$=$\delta_{ni}$=10$^3$ Hz. 
For the magnetic field strength of $B=$10 G, the ratio ${{Q_{ei}}/{Q_{in}}}$ is approximately 
$0.0023$, which means that the heating due to ion-neutral collisions is much stronger than the Ohmic heating in chromospheric jets. 

\cite{2004A&A...422.1073K} showed that viscosity effects are much smaller those 
of ion-neutral collisions 
in the solar chromosphere. Indeed, the ratio of the third (associated to viscous heating) and the second terms on the right hand-side of Equation~8 is approximately  

\begin{equation}
{{Q_{visc}}\over {Q_{in}}}\sim {{V^2 \beta}\over V^2_A}{{\delta_{in}\delta_{ii}}\over {\nu^2_{ci}}}, 
\label{rat2}
\end{equation}
where $\beta=8\pi p/B^2$ is the plasma beta parameter, $\delta_{ii}$ the ion-ion collision frequency, 
$\omega_{ci}$ the ion gyrofrequency, $V$ the plasma velocity, and $V_A$  the Alfv\'en speed. 
For an ion number density of 10$^{11}$ cm$^{-3}$ and temperature of 10$^4$ K, 
typical for chromospheric jets, the ion-ion collision frequency is $\delta_{ii}\approx$10$^5$ Hz. 
A velocity of $\mathrm{V=10~km~s^{-1}}$, magnetic field strength of 10 G, 
ion gyrofrequency $\nu_{ci}=10^5$~Hz, 
typical chromospheric Alfv\'en speed of $\mathrm{50~km~s^{-1}}$,  $\delta_{in}$=10$^3$ Hz  
and plasma beta $\beta=0.1$, 
gives a value for the ratio of viscous to ion-neutral heatings ${{Q_{visc}}/{Q_{in}}}$ 
$\sim$ 10$^{-5}$. Therefore, viscous heating is negligible compared to that by ion-neutral collisions. 

These estimates suggest that ion-neutral collisions (the second term on right-hand side of Equation 8) define the heating time of KH vortices. The heating time for certain harmonics can be written as  
\begin{equation}
t_{heat}\sim {{\beta \delta_{in} D^2}\over {V^2_A }} {{1-\xi_n}\over {\xi^2_n}}, 
\label{heat}
\end{equation}
where 
$D$ is the characteristic width of a KH vortex. 
If the  jet radius is $\mathrm{a=130~km}$, then the characteristic widths of $m=20-50$ harmonics are $D=2\pi a/m \approx 16-40$ km. 
Furthermore, the typical Alfv\'en speed of $\mathrm{50~km~s^{-1}}$, $\xi_n=0.5$, $\delta_{in}$=10$^3$ Hz and plasma beta $\beta=0.1$ 
yield a heating time of $\mathrm{\sim20-130~s}$. This simple estimation shows that the KH vortices may heat the plasma over timescales which are comparable to the lifetime of chromospheric jets.    

\begin{figure}[t]
\begin{center}
\includegraphics[width=8.7cm]{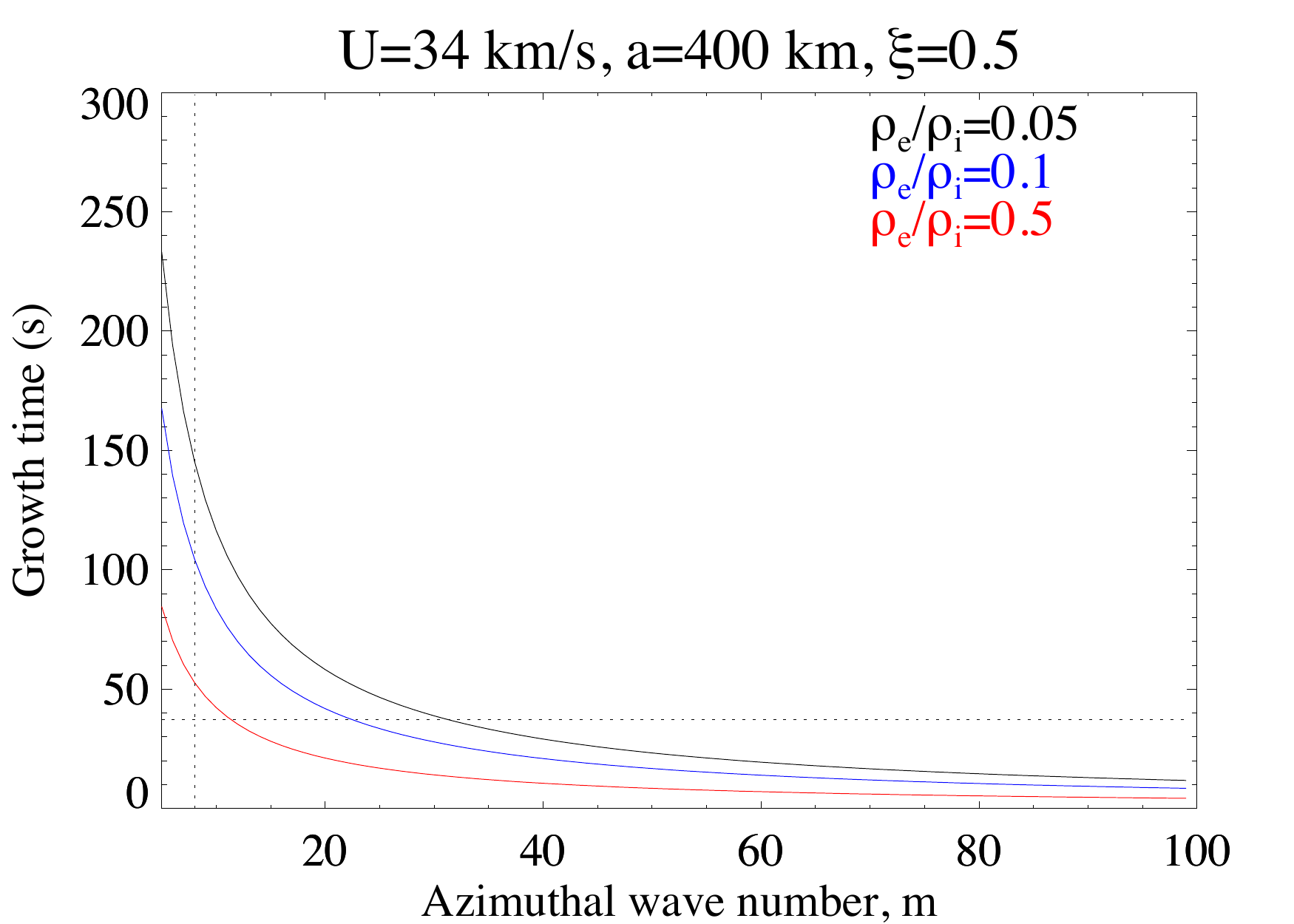}
\end{center}
\caption{KHI growth times as a function of $m$ for the large-scale twisted jets. The horizontal and vertical dotted lines indicate an vortex 
formation time ($\sim$37~s), and the number of azimuthal nodes ($\mathrm{m\approx8}$), respectively, estimated from Figure~\ref{fig6}.}
\label{fig7}
%\label{fig5}
\end{figure}

%%%%%%%%%%%%%%%%%%%%%%%%%%%%%%%%%%%%%%%%%%%%%%%%%%%%%%%%%%%%%%%%%%%%%%%%%%%%%%%%%%%%%%%%%%%%%%%%%%%%%%%%

\section{Discussion and Conclusion}

Some recent theoretical and observational studies suggest that  
H$\alpha$ jets, observed ubiquitously in the quiet Sun chromosphere, 
could be unstable to KHI due to the presence of velocity discontinuities between the surface of the jet and the surrounding plasma  
\citep{2011AIPC.1356..106Z, 2015ApJ...813..123Z}.  
Chromospheric jets, observed normally near network boundaries 
rooted in regions with photospheric magnetic bright points, 
are traditionally interpreted as overdense magnetic flux tubes with field aligned flows.  
However, it is suggested that at least some of the energetic, short-lived, type II spicular features 
may correspond to warps in two-dimensional sheets instead of flux tubes \citep{2011ApJ...730L...4J, 2012ApJ...755L..11J}. 
Chromospheric jets also exhibit transverse and torsional motions during their lifetime \citep{2012ApJ...752L..12D, 2014Sci...346D.315D}. 
The latter can produce a magnetic twist in the flux tubes where structures are formed.

To investigate theoretical aspects of the stability of chromospheric jets, we employ 
the KHI theory for the twisted jets developed in the incompressible ideal MHD limit. 
The chromospheric jets are modeled as cylindrical, high-density, twisted magnetic flux tubes moving in an axial direction in the twisted magnetized environment.
We derive dispersion equation governing the dynamics of twisted jets and solve them analytically in the large azimuthal wavelength limit.  
The solutions of the dispersion equation indicate that the perturbations with large azimuthal wave number $m$ are unstable to KHI with any upflow speed, $U$. %, in the long wavelength approximation and for the high order of $m$.
From the imaginary part of the solution and measured jet parameters we estimate the growth time of KHI (Figure~\ref{fig3}).
It shows that for REs, the growth times for 
$5\leqslant m\leqslant 40$ are between $\mathrm{5-250~s}$, 
and for $m\geqslant40$ the growth times are $\mathrm{1-35~s}$, depending on the flow speed and density ratio.

The turbulent motion and non-thermal heating produced by KHI could be observed as broadening of the spectral profiles of chromospheric jets. 
We employ H$\alpha$ spectra of the detected features to compute their Doppler widths.  
The analysis shows that the widths of the H$\alpha$ line profiles are broadened
with respect to the neighboring areas (Figure~4, 5). 
Similar findings have been reported in previous studies as well \citep{2009A&A...503..577C,2009ApJ...705..272R,2013ApJ...764..164S,2014ApJ...785..109L}. The analyzed REs have enhanced line widths all along the structure's lengths, 
indicating that if the line broadening is produced by turbulent motions associated with KHI vortices, 
then the longitudinal scale of the KHI should be much shorter than the length of the structures. 
Indeed, the estimated wavelengths of the KHI perturbation ($\mathrm{\lambda \sim 70 - 270~km}$) are much less than the REs measured lengths ($\mathrm{\sim3.5~Mm}$).

It must be noted that the approach used here to compute the Doppler width has certain limitations, as it attributes the 
changes in the absorption line profiles intensities to the line width and LOS velocities only, while  
intensity changes are in general proportional to density, wavelength dependent opacity, and the source function. 
Recently, \cite{2015ApJ...813..125K} has shown that the intensity changes in the H$\alpha$ line profile 
are highly dependent on the velocity gradient in the solar chromosphere, as it can create differences in the opacity between the red and blue wings of line core.
Furthermore, 
the H$\alpha$ line positions in our observations had a maximum step of 0.363 {\AA} which 
generates an uncertainty in our measurements of the Doppler parameters of the order of 7~\ks.  
Advanced cloud modeling with chromospheric radiative transfer calculations, together with observations at improved spectral resolution, 
would allow us to compute velocities and line widths for the chromospheric jets with higher accuracy.

Direct observational evidence of KHI perturbations in REs would be the detections of vortex-like features/azimuthal nodes at their boundaries.
The widths of the REs analyzed in this paper are only about twice as large as the spatial resolution at H$\alpha$, which
makes the KHI vortices/azimuthal nodes unresolved in the current dataset. 
However, as well as small-scale REs, we have detected a larger H$\alpha$ jet, 
which appears as a circular shaped absorption feature in the blue wing of H$\alpha$ (Figure~\ref{fig6}). 
Its radius ($\mathrm{\sim400~km}$) and LOS velocity 
($\mathrm{\sim34~km~s^{-1}}$) suggest that this jet could be the on-disk counterpart of 
a macrospicule or H$\alpha$ surge frequently observed at the solar limb. The morphology of the jet suggests that the plasma flow was oriented along the LOS.
Figure~\ref{fig6} shows that the structure develops azimuthal nodes over timescales of tens of seconds around its boundary. 
From a visual inspection we estimate $\mathrm{m\sim8}$ projected vortex-likes flows at around $\mathrm{t=37.52~s}$ (Figure~\ref{fig6}).    
Translating this number of observed azimuthal nodes into growth times using our theoretical model, we obtain $\mathrm{T\sim 52 - 145~s}$ for the density ratios of 0.5 to 0.05,  
which appears to be consistent with the timescale ($\sim$37 s) for the structure to develop the nodes in the image sequence presented in Figure~\ref{fig6}. 

KHI can produce the fast heating of chromospheric jets 
if the large-scale KHI vortices are decomposed
through nonlinear cascade of energy transfer to small scales. 
Furthermore, there are several damping mechanisms that could be responsible for a direct energy transfer of large-scale 
KHI vortices into heat without a non-linear cascade to the small scales. 
In the theoretical model presented in section 2 we used adiabatic approximation in an ideal regime for a fully ionized plasma, 
which neglects heating processes and non-ideal effects such as diffusivity, viscosity, radiative cooling, thermal conduction and Ohmic dissipation.
Therefore, the model presented cannot reproduce heating as a direct effect of KHI.
We have shown that the heating timescales associated with diffusivity, thermal conduction and radiative losses are much longer than the dynamic timescales of observed chromospheric jets. 
To investigate further non-linear, non-adiabatic effects we provide estimates of characteristic heating times for
other dissipation processes 
through the analysis of the energy equation in the partially ionized plasma. 
The results indicate that the ion-neutral collisions could be the most important process for the heating of the KH vortices and 
consequently the structure itself. For REs we estimate timescales of the heating due to the ion-neutral collisions, 
for KHI vortices with a high order of azimuthal wavenumbers, to be $\mathrm{\sim20-130~s}$, 
comparable to the lifetimes of these chromospheric jets.
We note that a similar approach has been used in the classical work of \cite{1965RvPP....1..205B}, where 
MHD wave equations are solved without neutrals and the results are used in an energy equation with neutrals 
 to study the damping of MHD waves and heating rate due to the ion-neutral collisions.
Such an approach allows a 
preliminary study of the possible KHI-associated heating in the cylindrical, twisted magnetic jets, for 
which there is no 
direct analytical solution when including partially ionized plasma in the basic MHD model presented in Section 2.

Due to the striking match between observed time-scales and those of ion-neutral heating at KHI vortices, further assessement of the importance of KHI to the heating of chromospheric jets, via 
detailed numerical simulations/forward modeling, is recommended.  
Furthermore, higher temporal, spatial and spectral resolution observations (e.g., GREGOR, SST, DKIST) will provide a better 
opportunity to resolve and study the KHI perturbations in chromospheric fine-scale structures. 

%%%%%%%%%%%%%%%%%%%%%%%%%%%%%%%%%%%%%%%%%%%%%%%%%%%%%%%%%%%%%%%%%%%%%%%%%%%%%%%%%%%%%%%%%%%%%%%%%%%%%%%%

\begin{acknowledgements}
The research leading to these results has received funding from the European Community's Seventh Framework Programme (FP7/2007-2013) under grant agreement no. 606862 (F-CHROMA).
The work of TVZ was supported by the Austrian "Fonds zur F\"{o}rderung der wissenschaftlichen Forschung under projects P26181-N27 and P25640-N27, and by FP7-PEOPLE-2010-IRSES-269299 project- SOLSPANET.
\end{acknowledgements}

%%%%%%%%%%%%%%%%%%%%%%%%%%%%%%%%%%%%%%%%%%%%%%%%%%%%%%%%%%%%%%%%%%%%%%%%%%%%%%%%%%%%%%%%%%%%%%%%%%%%%%%%

\bibliography{bibtex.bib}
\end{document}